\documentclass[10 pt, letter]{article}
\usepackage{graphicx}
\usepackage{caption}
\usepackage{amsmath}
\usepackage{amsthm}
\usepackage{enumerate}
\usepackage{subfig}
\usepackage{float}
\usepackage{amsfonts}
\usepackage{amssymb}
\usepackage{anysize}
\usepackage{helvet}
\usepackage{setspace}
\usepackage{multirow}
\usepackage{tabularx}
\usepackage{tabulary}
\usepackage{longtable}
\usepackage{float}
\usepackage{mathtools}
\title{\bf On Maxwell-like equations for Gravitational Field}
\author{Gustavo V. L\'opez\footnote{gulopez@udgserv.cencar.udg.mx}~\\ \\
 Departamento de F\'{i}sica, Universidad de Guadalajara,\\
 Blvd. Marcelino Garc\'{i}a Barragan y Calzada Ol\'{i}mpica, \\ 44200 Guadalajara, Jalisco, Mexico}
\begin{document}
\maketitle

\begin{abstract}
\noindent
For explicitly time depending mass density, which satisfies a continuity equation, it is shown that Maxwell-like equations for gravitational field follow naturally  without any need of General Relativity Theory approximation or related assumptions. As a consequences, it is shown that several features already known in Electrodynamics (Poynting vector, density of energy, tensor stress, radiation) are totally reproduced for gravitational field.
\end{abstract}
\newpage
\section{ Introduction}
Gravitational field is one of the most important fields created by Nature which has ( so far) pure attractive effect among anything with mass or energy[1,2], and its associated force is radial and inverse proportional to the square of the separation of the objects. For an arbitrary mass density distribution which is not depending explicitly on time, it is known that the gravitational field in the space can be found by solving  Poisson equation with proper boundary conditions [3]. However, Einstein [4] saw that gravity can be seen as the component of a pseudo-metric defined in a space-time manifold, where the shape of this manifold is due to  the mass and energy contained in it [5]. This was called the General Relativity Theory and have had enormous success explaining and predicting many gravitational phenomena in Nature [6]. However, this classical theory has had serious difficulties to construct a more fundamental theory (as many thinks) which is related with the quantization of gravity (the union of General Relativity with Quantum Mechanics [7]), where a graviton appears as a quantum particle associated to the interaction with gravity which has spin two [8]. This direction of research has its own value, and there are many scientists working on this topic. \\ \\
In addition, a big effort has been made to unify gravity with electrodynamics [8,9], but still a consistent mathematical and geometrical theory is missing. Alternatively, using perturbation theory on GR [10,11], or the decomposition of the Weyl and Maxwell tensors in electric and magnetic parts [12,19]. The so called Gravito-Electromagnetism (GEM) emerges as a result of these approaches, which is the expression of the equations to describe the gravitational field as a Maxwell-like type of equations. The similarity of gravity and electrodynamics equations comes since long time ago when people realized that, in the static case, both fields satisfy Poisson's equations, In fact, Heaviside [20] was one of the first persons who saw the similarities between Electromagnetism and Gravity through Poisson's equation, and in addition [21], for a current of mass motion as continuous fluid (${\bf J}=\rho{\bf v}$), he postulated a gravity field ${\bf h}$ having a curl analog to magnetic field in Maxwell's equations for Electrodynamics, in order to get a Maxwell-like equations for gravity, which, of course lead us to gravitational waves. Although his approach lack of mathematical justification, it can not be considered as a deduction, and it is done for a very particular case (fluid), the idea settled there was that Maxwell-like equations for gravity maybe could be arisen without GR theory. In a similar  axiomatic form, Yaroslav [25] has recently show the deduction of Maxwell-like equations for gravity without GR, where perihelion of Mercury and bending of light by massive object are presented. The idea of having a magnetic-like gravitational field in GR was reinforced  due to Lense-Thirring effect [22], where the angular momentum of a rotating body can be interpreted as a gravitational magnetic field, at large distances. Then, one can have the conclusion that GR is needed in order to see a gravitational field like GEM.  \\ \\
\noindent
On this paper, one will be focused  on other classical aspect of the gravitational field, to show that a full Maxwell-like equations for gravitational field  arises as a consequence of the continuity equation for the mass density and the usual Gauss' theorem.
\vskip1pc
\section{Maxwell-like equation for gravity}
For a time explicitly depending density of mass, $\rho({\bf x},t)$, where the following continuity equation is satisfied
\begin{equation}\label{cq}
\frac{\partial\rho}{\partial t}+\nabla\cdot{\bf J}=0,
\end{equation}
with ${\bf J}$ being the mass current density, which for a fluid-like system is given by ${\bf J}=\rho{\bf v}$, being ${\bf v}$  the fluid velocity. This continuity equation brings about the total mass conservation of the system, $M=\int_{\Re^3}\rho({\bf x},t)~d^3{\bf x}$.  Now, it is well known [2] that if the law of forces between to mass elements is of the form ${\bf F}\sim({\bf x_2-x_1})/|{\bf x_2-x_1}|^3$, where ${\bf x_1}$ and ${\bf x_2}$ are the vector position of the two elements, and $|{\bf x_2-x_1}| $ is their distance separation, then it follows that the Gauss' law for the gravitational field ${\bf R}$ in a compact region of the space $\Omega\in\Re^3$ is
\begin{equation}
\nabla\cdot{\bf R}=\begin{cases} 4\pi G\rho({\bf x},t)& \hbox{if}\quad\rho\in\Omega\\ 0&\hbox{if}\quad\rho\not\in\Omega\end{cases}
\end{equation}
where $G$ is the constant of gravitation ($G=6.674\times 10^{11}m^3/Kg\cdot s^2$). Differentiating this expression with respect the time, and using (\ref{cq}) in the region $\Omega$, one has
\begin{equation*}
 \nabla\cdot\frac{\partial{\bf R}}{\partial t}=4\pi G\frac{\partial\rho}{\partial t}=-4\pi G \nabla\cdot{\bf J},
 \end{equation*}
 or
 \begin{equation}
 \nabla\cdot\left(\frac{\partial{\bf R}}{\partial t}+4\pi G{\bf J}\right)=0.
 \end{equation}
 This equation implies that there must exist a vector field ${\bf W}$ such that
 \begin{equation}\label{cW}
 \frac{\partial{\bf R}}{\partial t}+4\pi G{\bf J}=\nabla\times{\bf W}.
 \end{equation}
Of course, there is a well known indetermination of this expression that can be absorbed in the definition of ${\bf R}$ or ${\bf W}$. It is necessary to mention that up to know non experiment has been made to try to find
the ${\bf W}$ gravitational field (something similar when one has a current in a wire and see what happen to the motion of a charge nearby), and this is the gravitational field so called ${\bf h}$  by Heaviside [21], defined on purpose to obtain Maxwell-like equations within his approach.   
So far, one knows $\nabla\cdot {\bf R}$ and $\nabla\times{\bf W}$ of the vector fields ${\bf R}$ and ${\bf W}$. To know $\nabla \times{\bf R}$ and $\nabla\cdot{\bf W}$, let us take the curl of (\ref{cW}) and use the known expression $\nabla\times(\nabla\times{\bf a})=\nabla(\nabla\cdot{\bf a})-\nabla^2{\bf a}$ for any vector field ${\bf a}$,
\begin{equation}\label{cRW}
\nabla(\nabla\cdot{\bf W})-\nabla^2{\bf W}=4\pi G\nabla\times{\bf J}+\frac{\partial}{\partial t}\bigl(\nabla\times{\bf R}\bigr).
\end{equation}
To decouple equations for ${\bf R}$ and ${\bf W}$, one can choose the following relations
\begin{subequations}
\begin{equation}\label{qq1}
\nabla\cdot{\bf W}=\alpha \quad\hbox{(constant)}
\end{equation}
and
\begin{equation}\label{qq2}
\nabla\times{\bf R}\sim Ð\frac{\partial{\bf W}}{\partial t}.
\end{equation}
\end{subequations}
Denoting the constant of proportionality of (\ref{qq2}) as $1/\lambda^2$, one has for (\ref{cRW}) the following expression
\begin{equation}\label{w1}
\nabla^2{\bf W}-\frac{1}{\lambda^2}\frac{\partial^2{\bf W}}{\partial t^2}=-4\pi G\nabla\times{\bf J}.
\end{equation} 
Now, taking the curl of (\ref{qq2}) , it follows that
\begin{equation*}
\nabla\times(\nabla\times{\bf R})=-\frac{1}{\lambda^2}\frac{\partial \nabla\times{\bf W}}{\partial t}
\end{equation*}
or
\begin{equation}
\nabla (\nabla\cdot{\bf R})-\nabla^2{\bf R}=-\frac{1}{\lambda^2}\frac{\partial}{\partial t}\left(\frac{\partial{\bf R}}{\partial t}+4\pi G{\bf J}\right),
\end{equation}
which can be written as
\begin{equation}\label{r1}
\nabla^2{\bf R}-\frac{1}{\lambda^2}\frac{\partial^2{\bf R}}{\partial t^2}=\frac{4\pi G}{\lambda^2}\frac{\partial {\bf J}}{\partial t}+4\pi G\nabla\rho.
\end{equation}
Equations (\ref{w1}) and (\ref{r1}) represent non homogeneous wave equations with known sources. The wave propagates with speed "$\lambda$", and by selecting $\alpha=0$ on (\ref{qq1}), the equations which determine the gravitational fields ${\bf R}$ and ${\bf W}$ have a Maxwell-like form
\begin{eqnarray}
&\nabla\cdot{\bf R}=4\pi G\rho({\bf x},t)\\ \nonumber\\
&\displaystyle\nabla\times {\bf R}=-\frac{1}{\lambda^2}\frac{\partial{\bf W}}{\partial t}\label{s2}\\ \nonumber\\
&\displaystyle\nabla\cdot{\bf W}=0\\ \nonumber\\
&\displaystyle\nabla\times{\bf W}=4\pi G{\bf J}({\bf x},t)+\frac{\partial{\bf R}}{\partial t}\label{c4}
\end{eqnarray}
Thus, one may say in summary that if the Gauss' theorem is valid for the gravitational field ${\bf R}$, and the continuity equation is  satisfied for $\rho$ and ${\bf J}$, then, there must exist a gravitational vector field ${\bf W}$ such that ${\bf W}$ and ${\bf R}$ satisfy inhomogeneous  wave equations which propagate with an unknown speed "$\lambda$." One must mention that there has not been a direct measurement of the speed of gravitational waves, but it is assume that gravity also travel to the speed of light [6].\\Ê\\
Now, in terms of the vector and scalar potentials, ${\bf A}_W$ and $\Phi_W$, which is well known where they come from, the gravity fields can be written as
\begin{subequations}
\begin{equation}
{\bf R}=-\nabla\Phi_W-\frac{1}{\lambda^2}\frac{\partial{\bf A}_W}{\partial t}
\end{equation}
and
\begin{equation}
{\bf W}=\nabla\times{\bf A}_W.
\end{equation}
\end{subequations}
Of course, and identical relation with the electrodynamics Maxwell's equation can be gotten by choosing (\ref{s2}) and (\ref{c4}) as
\begin{subequations}
\begin{equation}
\nabla\times{\bf W}=\frac{4\pi G}{\lambda}{\bf J}+\frac{1}{\lambda}\frac{\partial{\bf R}}{\partial t}
\end{equation}
and
\begin{equation}
\nabla\times{\bf R}=-\frac{1}{\lambda}\frac{\partial{\bf W}}{\partial t}.
\end{equation}
\end{subequations}
If one applies  the same initial hypothesis to the electric field in the case of electrodynamics, one see that existence of the magnetic field and the displacement current must appear as a direct consequence of the continuity equation for the density of charge and current. The decouple of the resulting equations would bring about the usual Maxwell's equations and the nonhomogeneous wave equations for the electric and magnetic fields, with the waves traveling at the speed of light.
\section{Lorentz-like Force}
 Firstly, one needs to say that in the MKS system of units, one has the following units of our quantities and vectors
 \begin{equation}
 [\rho]=Kg/m^3,\quad [{\bf J}]=Kg/m^2\cdot s,\quad [\lambda]=m/s,\quad [{\bf R}]=m/s^2,\quad [{\bf W}]=m^2/s^3.
 \end{equation} 
 Now, given the gravitational field ${\bf R}({\bf x},t)$ defined in the space-time, the force felt by and object of mass $M$ located at the point ${\bf x}$  at the time "$t$" due to the field ${\bf R}$ is given by 
 \begin{equation}
{\bf F}_R=M{\bf R}({\bf x},t),
\end{equation}
which is well known. However, given the field ${\bf W}({\bf x},t)$ in the space-time, the only experimental evidence that this field could be associated to a magnet-like force is the Lense-Thirring effect [ 18]. In this way, the force associated to the gravitational field ${\bf W}$ would be
\begin{equation}
{\bf F}_W=\frac{M}{\lambda^2}{\bf v}\times{\bf W}({\bf x},t).
\end{equation} 
If one considers that these are true, one gets a Lorentz-like force expression for an object of mass $M$ under the gravitational interaction with the fields ${\bf R}$ and ${\bf W}$,
\begin{equation}
{\bf F}=M{\bf R}({\bf x},t)+\frac{M}{\lambda^2}{\bf v}\times{\bf W}({\bf x},t),
\end{equation}
where ${\bf v}$ is the velocity of the object.
\section{Energy Density, Poynting Vector, Stress Matrix}
Let us write the Maxwell-like equations for the gravitational fields ${\bf R}$ and ${\bf W}$ (MKS units) as
\begin{eqnarray}
& &\nabla\cdot{\bf R}=4\pi G\rho\label{fa}\\
& &\nabla\times{\bf R}=-\frac{1}{\lambda}\frac{\partial{\bf W}}{\partial t}\label{fb}\\
& &\nabla\cdot{\bf W}=0\label{fc}\\
& &\nabla\times{\bf W}=\frac{4\pi G}{\lambda}{\bf J}+\frac{1}{\lambda}\frac{\partial{\bf R}}{\partial t}\label{fd},
\end{eqnarray}
where  the resulting decoupled equations for ${\bf R}$ and ${\bf W}$ are the inhomogeneous wave equations
\begin{eqnarray}
& &\nabla^2{\bf R}-\frac{1}{\lambda^2}\frac{\partial^2{\bf R}}{\partial t^2}=4\pi G(\nabla\rho)+\frac{4\pi G}{\lambda^2}\frac{\partial {\bf J}}{\partial t}\label{f3a}\\ \nonumber\\
&&\nabla^2{\bf W}-\frac{1}{\lambda^2}\frac{\partial^2{\bf W}}{\partial t^2}=-\frac{4\pi G}{\lambda}\nabla\times{\bf J}\label{f3b}
\end{eqnarray}
 and write the Lorentz-like law of force, acting over an object of mass $m$, as
\begin{equation}\label{for}
{\bf F}=m{\bf R}+\frac{m}{\lambda}{\bf v}\times{\bf W},
\end{equation}
where ${\bf v}$ is the velocity of the object. The units of the vector fields ${\bf R}$ and ${\bf W}$ are the same, $m/s^2$. As pointed out before, the field ${\bf W}$, the speed $\lambda$, and the force ${\bf F}=m{\bf v}\times{\bf W}/\lambda$ still requiere direct experimental verification. \\ \\
Now, let us proceed as one does in Electrodynamics Theory [24]. From (\ref{fc}) , one knows that there exists a vector field ${\bf A}$ (vector potential) such that
\begin{equation}\label{w1}
{\bf W}=\nabla\times{\bf A}.
 \end{equation} 
 Substituting this in (\ref{fb}), it follows that there exists a function $\Phi$ (scalar potential) such that
 \begin{equation}\label{r1}
 {\bf R}=-\nabla\Phi-\frac{1}{\lambda}\frac{\partial {\bf A}}{\partial t}.
 \end{equation}
Using (\ref{w1}) and (\ref{r1})  in the expressions (\ref{fa}) and (\ref{fd}), and choosing the Lorentz-like gauge
\begin{equation}\label{ge}
\nabla\cdot{\bf A}+\frac{1}{\lambda}\frac{\partial\Phi}{\partial t}=0,
\end{equation}
one gets decoupled equations for the potentials $\Phi$ and ${\bf A}$  which also satisfy  inhomogeneous wave equations
\begin{subequations}
\begin{equation}\label{wa1}
\nabla^2\Phi-\frac{1}{\lambda^2}\frac{\partial^2\Phi}{\partial t^2}=-4\pi G\rho
\end{equation}
and
\begin{equation}\label{wa2}
\nabla^2{\bf A}-\frac{1}{\lambda^2}\frac{\partial^2{\bf A}}{\partial t^2}=-\frac{4\pi G}{\lambda}{\bf J}.
\end{equation}
\end{subequations}
The power density is related with the mechanical energy density as
\begin{equation}
P=\frac{\partial u}{\partial t}={\bf J}\cdot{\bf R},
\end{equation}
which can be written using (\ref{fd}) and (\ref{fb}) as
\begin{equation}
\frac{\partial u}{\partial t}=-\frac{\lambda}{4\pi G}\biggl(\nabla\times{\bf W}-\frac{1}{\lambda}\frac{\partial{\bf r}}{\partial t}\biggr)\cdot{\bf R}-\frac{\lambda}{4\pi G}\biggl(\nabla\times{\bf r}+\frac{1}{\lambda}\frac{\partial {\bf W}}{\partial t}\biggr)\cdot{\bf W},
\end{equation}
bringing about the relation
\begin{equation}\label{ss1}
\frac{\partial u}{\partial t}=-\nabla\cdot{\bf S}-\frac{\partial u_g}{\partial t},
\end{equation}
where ${\bf S}$ and $u_g$ are the Poynting vector and the gravitational energy density,
\begin{equation}
{\bf S}=\frac{\lambda}{4\pi G}\bigl({\bf R}\times{\bf W}\bigr)
\end{equation}
and
\begin{equation}
u_g=\frac{1}{8\pi G}\biggl(|{\bf R}|^2+|{\bf W}|^2\biggr).
\end{equation}
As one can see, the concept of gravitational energy is well defined, and the Poynting vector ${\bf S}$ has the usual meaning of flux of density energy per unit area per unit time. For pure gravitational field ($u=0$), one obtains the continuity equation
\begin{equation}\label{css}
\nabla\cdot{\bf S}+\frac{\partial u_g}{\partial t}=0,
 \end{equation} 
implying that the gravitational energy velocity is
\begin{equation}\label{vs}
{\bf v}_g=\frac{\bf S}{u_g}=2\lambda\frac{{\bf R}\times{\bf W}}{|{\bf R}|^2+|{\bf W}|^2}.
\end{equation}
In vacuum ($\rho=0$ and ${\bf J}={\bf 0}$), a solution of (\ref{f3a}) and (\ref{f3b}) are plane gravitational waves, ${\bf R}, {\bf W}\sim e^{i({\bf k}\cdot{\bf x}-\omega t)}$, where ${\bf k}$ and $\omega$ are the vector wave number and frequency of gravitational wave, and one has  the properties: ${\bf R}\cdot{\bf k}=0$, ${\bf W}\cdot{\bf k}=0$, and  ${\bf W}=\widehat{\bf k}\times{\bf R}$. From  (\ref{vs}) it follows that the velocity of the gravitational plane wave is  ${\bf v}_g=\lambda {\bf k}$.\\ \\
Now, assuming the current density of mass as a fluid in motion, where every element of the fluid is moving with velocity ${\bf v}$,  the current density is determined by ${\bf J}=\rho{\bf v}$, and from the expression (\ref{for}), the density of force at any point on the fluid is 
\begin{equation}
{\bf f}=\frac{d{\bf F}}{dV}=\rho{\bf R}+\frac{1}{\lambda}{\bf J}\times{\bf W},
\end{equation}
This expression can be written using (\ref{fa}) and (\ref{fd}) as
\begin{equation}
{\bf f}=\frac{\partial{\bf P}_m}{\partial t}=\frac{1}{4\pi G}\bigl(\nabla\cdot{\bf R}\bigr){\bf R}+\frac{1}{4\pi G}\bigl(\nabla\times{\bf W}-\frac{1}{\lambda}\frac{\partial{\bf R}}{\partial t}\bigr)\times{\bf W},
\end{equation}
where ${\bf P}_m$ is the density of mechanical linear momentum of the system. After adding the zero parts $(\nabla\cdot {\bf W})\cdot {\bf R}$ and $(\nabla\times{\bf R}+\partial{\bf W}/\partial \lambda t)\times {\bf R}$ and doing some manipulations, this expression can be written as
\begin{equation}
\frac{\partial{\bf P}_m}{\partial t}=\nabla\cdot T-\frac{\partial{\bf P}_g}{\partial t},
\end{equation}
where ${\bf P}_g$ and $T$ are the gravitational linear momentum and tensor stress,
\begin{subequations}
\begin{equation}
{\bf P}_g=\frac{1}{\lambda^2}{\bf S}=\frac{1}{4\pi G\lambda}\bigl({\bf R}\times{\bf W}\bigr)
\end{equation}
and
\begin{equation}
T=\frac{1}{4\pi G}\biggl\{{\bf R}\otimes{\bf R}+{\bf W}\otimes{\bf W}\biggr\}-u_gI,
\end{equation}
\end{subequations}
being $\otimes$ the tensorial product, and I being the $3\times 3$-identity matrix. So, the gravitational field has a well defined linear momentum and stress tensor. 
\section{Formulation in space-time $\Re^4$}
Consider the following vectors in $\Re^4$ with components
\begin{equation}
x_{\mu}: ({\bf x},\lambda t),\quad\quad x^{\mu}: ({\bf x},-\lambda t)
\end{equation}
Consider also the following vector fields defined on $\Re^4$ (space-time) with components
\begin{equation}
j_{\mu}: ({\bf J}, \lambda\rho),\quad\quad A_{\mu}: ({\bf A}, \Phi),\quad\quad S_{\mu}: ({\bf S}, \lambda u_g)
\end{equation}
and the $4\times 4$ anti-symmetric matrix defined by the following matrix elements
\begin{equation}
F_{\mu\nu}=\frac{\partial A_{\mu}}{\partial x_{\nu}}-\frac{\partial A_{\nu}}{\partial x_{\mu}},
\end{equation}
or, from the expressions (\ref{w1}) and (\ref{r1}), this matrix is explicitly given by
\begin{equation}
F=\begin{pmatrix}0&W_z&-W_y&R_x\\ -W_z&0&W_x&R_y\\ W_y&-W_x&0&R_z\\-R_x&-R_y&-R_z&0\end{pmatrix}.
\end{equation}
Using the operators
\begin{equation}
\partial_{\mu}: \left(\nabla,-\frac{\partial}{\partial(\lambda t)}\right),\quad\quad \partial^{\mu}: \left(\nabla, \frac{\partial}{\partial(\lambda t)}\right) 
\end{equation}
and the operator (assuming Einstein summation convention)
\begin{equation}
\square\equiv\partial^{\mu}\partial_{\mu}=\nabla^2-\frac{1}{\lambda^2}\frac{\partial^2}{\partial t^2},
\end{equation}
one can have the following identifications
\begin{eqnarray}
& & \nabla\cdot{\bf J}+\frac{\partial\rho}{\partial t}=0\longrightarrow \partial^{\mu}j_{\mu}=0\label{id1}\\ 
& &\nabla\cdot{\bf A}+\frac{1}{\lambda}\frac{\partial\Phi}{\partial t}=0\longrightarrow \partial^{\mu}A_{\mu}=0\label{id2}\\
& &\nabla\cdot{\bf S}+\frac{\partial u_g}{\partial t}=0\longrightarrow \partial^{\mu}S_{\mu}=0\label{id3}\\ \nonumber \\
& &\begin{cases} \nabla\cdot{\bf R}=4\pi G\rho\\ \\ \displaystyle\nabla\times{\bf W}-\frac{1}{\lambda}\frac{\partial{\bf R}}{\partial t}=\frac{4\pi{\bf J}}{\lambda}\end{cases}\longrightarrow \partial^{\mu}F_{\mu\nu}=\frac{4\pi G}{\lambda}j_{\nu}\label{id4}\\ \nonumber \\ \\
& &\begin{cases}\nabla\cdot{\bf W}=0\\ \\ \displaystyle\nabla\times{\bf R}+\frac{1}{\lambda}\frac{\partial{\bf W}}{\partial t}=0\end{cases}\longrightarrow \partial^{\alpha}F_{\mu\nu}+\partial^{\mu}F_{\nu\alpha}+\partial^{\nu}F_{\alpha\mu}=0\label{id5}
\end{eqnarray}
In fact, the last expression follows necessarily due to anti-symmetric definition of $F$. Wave equations (\ref{f3a}), (\ref{f3b}), (\ref{wa1}) and (\ref{wa2}) can be written as
\begin{equation}
\square F_{\mu\nu}=-\frac{4\pi G}{\lambda}(\partial_{\mu}j_{\nu}-\partial_{\nu}j_{\mu}\label{ip1})
\end{equation}
and
\begin{equation}
\square A_{\mu}=-\frac{4\pi G}{\lambda}j_{\mu}\label{ip2}.
\end{equation}
This formulation is convenient to see the change of form of these expression under a change of inertial reference system.  
\section{Change of Inertial Reference System}
Let $S$ and $S'$ be two inertial reference systems which at $t'=t=0$ have a common origin, and $S'$ is moving with respect to $S$ with a constant velocity ${\bf V}$, the reference system $S$ is fixed with respect the far away Galaxies in our Universe. Let us  defined $\vec\beta$ as 
\begin{equation}
\vec\beta={\bf V}/\lambda,
\end{equation}
the normalized velocity of $S'$. Then, there must be a nonsingular matrix $\Lambda=\Lambda(\vec\beta)$ such that the space-time of the system $S$ can be transformed on the space-time of the system $S'$,
\begin{equation}
x_{\mu}'=\Lambda_{\mu}^{\alpha}(\vec\beta)x_{\alpha},
\end{equation}
such that
\begin{equation}
\Lambda(\vec\beta_2)\cdot\Lambda(\vec\beta_1)=\Lambda(\vec\beta_3),\quad \Lambda^{-1}(\vec\beta)=\Lambda(-\vec\beta),\quad \Lambda({\bf 0})=I.
\end{equation}
In other words, the set $\{\Lambda(\vec\beta)\}_{\beta\in[0,1]}$ has the structure of group. So, the transformation of the fields defined in $\Re^4$  that one would expect are of the form
\begin{eqnarray}
& & j_{\mu}'=\Lambda_{\mu}^{\alpha}(\vec\beta)j_{\alpha}\\
& &A_{\mu}'=\Lambda_{\mu}^{\alpha}(\vec\beta)A_{\alpha}\\
& &S_{\mu}'=\Lambda_{\mu}^{\alpha}(\vec\beta)S_{\alpha}\\
& & F'=\Lambda^{-1}(\vec\beta)\cdot F\cdot\Lambda(\vec\beta).
\end{eqnarray}
Of course, (\ref{id1}), (\ref{id2}), (\ref{id3}), (\ref{id4}), (\ref{id5}), (\ref{ip1}), and (\ref{ip2}) are invariant of form (they transform covariantly) under a change of an inertial reference system, although this transformation has not given explicitly. In addition, one has that 
\begin{equation}
tr F'^2=tr\biggl(\Lambda^{-1}(\vec\beta)\cdot F\cdot\Lambda(\vec\beta)\biggr)^2=tr F^2,
 \end{equation} 
 that is, $tr F^2$ is an invariant under these transformations. Therefore, one can define the Lagrangian density for the gravitational field as
 \begin{equation}\label{LaG}
 {\cal L}=\frac{1}{16\pi G}tr F^2+\frac{1}{\lambda}j_{\mu}A^{\mu},
  \end{equation} 
  and it is known from Electrodynamics [24] that Euler-Lagrange equations applied to this Lagrangian leads us to Maxwell-like equations (\ref{id4}) and (\ref{id5}).\\ \\
  This group of transformations also will leave some metric (defined in $\Re^4$) invariant. If this metric is given in the inertial reference system $S$ by
  \begin{equation}
  ds^2=d\xi^t\cdot g\cdot d\xi=\sum_{\mu\nu}g_{\mu\nu} dx^{\mu}\otimes dx^{\nu},
  \end{equation}
  where $\xi=({\bf x},\lambda t)$,  $\xi^t$ is the transpose vector, and $g$ is a $4\times 4$ real constant nonsingular matrix. In the inertial reference system $S'$, this metric is given by
  \begin{equation}
  ds'^2=d\xi'^t\cdot g\cdot d\xi', \quad \quad\hbox{with}\quad\ d\xi'=\Lambda^t(\vec\beta) d\xi .
  \end{equation}
  So, one will have invariance of the metric ($ds'^2=ds^2$) if it follows that
  \begin{equation}
  \Lambda(\vec\beta)\cdot g\cdot\Lambda^t(\vec\beta)=g
  \end{equation}
  Then, one chooses  as the reference systems $S'$ that one which goes with the object (${\bf x}'={\bf 0}$), and the time defined on this system (even it is not really an inertial system) is called ``proper time'' $\tau$. Thus, one gets
  \begin{equation}
  \lambda^2d\tau^2=d\xi^t\cdot g\cdot d\xi.
  \end{equation}
Factorizing $dt$ from the right hand side, one can obtain a relation between the proper time and the time measured in the system $S$ of the form
\begin{equation}\label{tau}
d\tau=\frac{1}{q(\widetilde\beta)}dt, \quad\hbox{or} \quad \frac{d}{d\tau}=q(\widetilde\beta)\frac{d}{dt},
\end{equation}
where $q(\widetilde\beta)$ is the resulting function of this factorization, and $\widetilde\beta$ is related with the velocity of the object (seen from the system $S$), $\widetilde\beta={\bf v}/\lambda$. The equations of motion invariant under the set of transformations $\{\Lambda(\vec\beta)\}_{0\le\beta\le 1}$ must be of similar form as in special theory of relativity [24],
\begin{equation}\label{req}
m\frac{d^2x^{\mu}}{d\tau^2}=f^{\mu},\quad\quad f^{\mu}: \biggl(q(\widetilde\beta){\bf F},\frac{q(\widetilde\beta)}{{\lambda}} \widetilde\beta\cdot {\bf F}\biggr).
\end{equation}
This formulation will be useful later on when radiation reaction force for gravitational field  be studied. For our purposes, it is not necessary to know the set of transformations neither the function $q(\widetilde\beta)$, if , however, one assumes that Poincaré-Lorentz-like transformations can be used here,  the velocity of the inertial system $S'$ is ${\bf V}=(0,0,V)$ ( with $\beta=V/\lambda$ ), and the normalized velocity of the object is $\widetilde\beta=(\tilde{\beta}_x,\tilde{\beta}_y,\tilde{\beta}_z)$  ( with ${\widetilde\beta}_i=v_i/\lambda$), one would have
\begin{equation}
\Lambda(\vec\beta)=\begin{pmatrix}1&0&0&0\\0&1&0&0\\0&0&\gamma&-\beta\gamma\\0&0&-\beta\gamma&\gamma\end{pmatrix}
,\quad\quad q(\widetilde\beta)=(1-\widetilde\beta^2)^{-1/2}
\end{equation}
\section{Gravitational Waves Emission}
It is known that the particular solution of the equations (\ref{wa1}) and (\ref{wa2}) are given by the convolution of the inhomogeneity with the fundamental solution of the wave equations [23],
\begin{equation}
\Phi({\bf x},t)=-4\pi G({\cal E}\star\rho),\quad\quad {\bf A}({\bf x},t)=-\frac{4\pi G}{\lambda}({\cal E}\star{\bf J}),
\end{equation}
where ${\cal E}$ is the fundamental solution,
\begin{equation}
{\cal E}({\bf x},t)=-\frac{1}{4\pi}\frac{\delta(t-|{\bf x}|/\lambda)}{|{\bf x}|}.
\end{equation}
This brings about the known retarded potentials
\begin{equation}\label{ret1}
\Phi({\bf x},t)=G\int_{\Omega\times\Re}\frac{\rho({\bf x}',t')\delta(t-t'-|{\bf x-x'}|/\lambda)}{|{\bf x-x'}|}d^3{\bf x}'dt'
\end{equation}
and
\begin{equation}
{\bf A}({\bf x},t)=\frac{G}{\lambda}\int_{\Omega\times\Re}\frac{{\bf J}({\bf x}',t')\delta(t-t'-|{\bf x-x'}|/\lambda)}{|{\bf x-x'}|}d^3{\bf x}'dt',
\end{equation}
where $\Omega\subset\Re^3$ is the domain where $\rho$ and ${\bf J}$ are defined. For a point object of mass $m$ which is moving arbitrarily, having the position ${\bf x}_m(t)$ and velocity ${\bf v}_m(t)$, 
with $\rho({\bf }',t')=m\delta\bigl({\bf x}'-{\bf x}_m(t')\bigr)$ and ${\bf J}({\bf x}',t')=m{\bf v}_m(t')\delta\bigl({\bf x}'-{\bf x}_m(t')\bigr)$, a Li\'enard-Wiechert potentials  are gotten, and the resulting gravitational fields are of the form
\begin{equation}\label{rrw}
{\bf R}={\bf R}_{\beta}+{\bf R}_{\dot\beta}, \quad\quad {\bf W}=\hat{\bf r}\times {\bf R},
\end{equation}
where ${\bf r}={\bf x}-{\bf x}'$ is the vector going from the object position ${\bf x}'$, to the observer position ${\bf x}$ , $\hat{\bf r}={\bf r}/|{\bf r}|$ is the unitary vector, and the gravitational fields ${\bf R}_{\beta}$ and ${\bf R}_{\dot\beta}$ are
\begin{subequations}
\begin{equation}\label{rb}
{\bf R}_{\beta}=\frac{Gm(\hat{\bf r}-\vec\beta)(1-\beta^2)}{r^2(1-\hat{\bf r}\cdot\vec\beta)^3}\biggr|_{t'=t-r/\lambda}
\end{equation}
and
\begin{equation}\label{rdb}
{\bf R}_{\dot\beta}=\frac{Gm}{\lambda}\frac{\hat{\bf r}\times\bigl[(\hat{\bf r}-\vec\beta)\times\dot{\vec\beta}\bigr]}{r(1-\hat{\bf r}\cdot\vec\beta)^3}\biggr|_{t'=t-r/\lambda},
\end{equation}
\end{subequations}
being $\vec\beta={\bf v}_m(t)/\lambda$ the normalized velocity of the object, and $t'$ is the retarded time. The gravitational power emitted by the accelerated object per solid angle is
\begin{equation}
\frac{dP}{d\Omega}=(1-\hat{\bf r}\cdot\vec\beta)r^2({\bf S}\cdot\hat{\bf r}),
 \end{equation} 
 and using (\ref{rrw}), (\ref{rb}) , and ({\ref{rdb}), one gets
 \begin{equation}\label{pow}
 \frac{dP}{d\Omega}=\frac{(Gm)^2}{4\pi\lambda}\frac{\bigl|\hat{\bf r}\times\bigl[(\hat{\bf r}-\vec\beta)\times\dot{\vec\beta}\bigr|^2}{(1-\hat{\bf r}\cdot\vec\beta)^5}
 \end{equation}
 Thus, any object of mass $m$ which is accelerated will emit gravitational radiation. Of course, the object must have a huge mass in order for this radiation to be observed. This result contrast a lot with the given by General Relativity since in this theory the object must have a quadrupole configuration in order to emits gravitational energy [26], and the gravitational waves emitted are of quadrupole type waves.\\Ê\\
 If one has a charged particle of charge $q$ and mass $m$ which is accelerated, this particle will emits electromagnetic [24]  and gravitational energy such that the ratio of electromagnetic to gravitational energy emitted is
 \begin{equation}
 \frac{(dP/d\Omega)_{\cal E}}{(dP/d\Omega)_g}=\left(\frac{q}{4\pi\epsilon_0 Gm}\right)^2.
 \end{equation}
 where $\epsilon_0=8,854\times 10^{-12}F/m$ is the constant dielectric of the vacuum. For the electron, $|q_e|=1.6\times 10^{-19} C$, this ratio is of the order of $10^{39}$, whatever be its acceleration, that is, the electromagnetic radiation totally dominate the gravitational radiation. In order for a object of mass $m$ and charge $q$ to emits the same gravitational energy as the electromagnetic energy emission, it would require that its mass would be  $m\approx q/4\pi\epsilon_0G$. For and object having an electron charge, this mass would be of the order of $21~Kg$. 
 \section{Radiation Reaction Force for Gravitational Field}
 In this section, one will follow the approach given in reference [27] for  the radiation reaction force in terms of the external force. An accelerated object  of mass $m$ dissipate energy due to gravitational waves emission that the object does during its accelerated motion. From the expressions (\ref{tau}) and (\ref{req}), the modified equation of motion due to to the transformation that leave invariant of form the Maxwell-like equations is
 \begin{equation}
 m\frac{d(q(\widetilde\beta){\bf v})}{dt}={\bf F},
  \end{equation} 
  where ${\bf F}$ is the external force, and $\widetilde\beta={\bf v}/\lambda$. Although the function $q(\widetilde\beta)$ is really unknown,  it is possible to assume the differentiation  and to write this expression as the Newtonian equation of motion
  \begin{equation}
m\frac{d\widetilde\beta}{dt}=\widetilde{\aleph}(\widetilde\beta){\bf F},  
  \end{equation}
where $\widetilde\aleph$ is some $3\times 3$ matrix function depending on $\widetilde\beta$. Using this expression in  
(\ref{pow}), it follows that
\begin{equation}
\frac{dP}{d\Omega}=\frac{(Gm)^2\bigl|\widetilde{\aleph}{\bf F}\bigr|^2}{4\pi\lambda}
\frac{\bigl|{\bf e}_{_1}\cos\phi_{_1}-\tilde\beta{\bf e}_{_2}\cos\phi_{–2}\bigr|^2}{(1-\tilde\beta\cos\theta)^5},
\end{equation}
where ${\bf e}_{_1}$ and $\phi_{_1}$ are the direction and the angle resulting from the expression $\hat{\bf r}\times(\hat{\bf r}\times\widetilde{\aleph}{\bf F})$,  ${\bf e}_{_2}$ and $\phi_{_2}$ are the direction and the angle resulting from the expression $\hat{\bf r}\times(\widetilde\beta\times\widetilde{\aleph}{\bf F})$, and $\theta$ is the angle between $\hat{\bf r}$ and $\widetilde\beta$.
Making the integration of this expression with respect the solid angle and over the time intervale within the external force is acting, $[0,t]$, one gets the energy dissipated by the accelerated object as
\begin{equation}\label{ed1}
U(t)=\frac{(Gm)^2}{4\pi\lambda}\int_0^t\big|\widetilde{\aleph}{\bf F}\bigr|^2dt \int_{\Omega}\frac{\bigl|{\bf e}_{_1}\cos\phi_{_1}-\tilde\beta{\bf e}_{_2}\cos\phi_{_2}\bigr|^2}{(1-\tilde\beta\cos\theta)^5}d\Omega.
\end{equation}
Now, one assumes that this energy is due to the work done by a nonconservative force ${\bf F}_{rad}$ which moves the object from the position ${\bf x}_0$ at $t=0$ to the position ${\bf x}$ at a time $t > 0$. Because $d{\bf x}={\bf v}~dt$, one obtains
\begin{equation}\label{ed2}
U(t)=\int_0^t{\bf F}_{rad}\cdot{\bf v}~dt.
\end{equation}
Equaling (\ref{ed1}) and (\ref{ed2}),  knowing that the angle between ${\bf F}_{rad}$ and ${\bf v}$ must be $\pi$  (dissipative direction), and having ${\bf F}_{rad}\sim \hat{\bf n}$ with $\hat{\bf n}={\bf v}/v$, one gets finally and expression for the radiation reaction force of gravitational field as
\begin{equation}\label{frr}
{\bf F}_{rad}=-{\bf v}\frac{(Gm)^2\bigl|\widetilde{\aleph}{\bf F}\bigr|^2}{4\pi\lambda v^2}\int_{\Omega}
\frac{\bigl|{\bf e}_{_1}\cos\phi_{_1}-\beta{\bf e}_{_2}\cos\phi_{_2}\bigr|^2}{1-\beta\cos\theta)^5}d\Omega,
\end{equation}
where $v$ is the magnitude of the velocity of the object.
Therefore, the dynamical equations of motion which take into account the energy lost by gravitational radiation due to the accelerated object is
\begin{equation}
m\frac{d\bigl(q(\widetilde\beta){\bf v}\bigr)}{dt}={\bf F}+{\bf F}_{rad},
\end{equation}
with ${\bf F}_{rad}$ given by (\ref{frr}). Note that whenever ${\bf F}={\bf 0}$, one has that ${\bf F}_{rad}={\bf 0}$ too. If the external force is zero, there is not acceleration, implying that there is not radiation of gravitational waves, one expects that this happen experimentally.  
\\Ê\\
Although there have been several indirect facts [29] that indicates that the gravitational waves speed could be the same speed of light ($\lambda=c=2.99792\times 10^8m/s$ [30]), it is necessary a direct measurement of this important   parameter for gravitational waves. The gravitational field ${\bf W}$ appears naturally from mathematical analysis, but requires experimental verification together with the law of force that it may define, ${\bf F}=m {\widetilde\beta}\times {\bf W}$. It looks like that the transformations $\{\Lambda(\vec\beta)\}_{0\le\beta\le 1}$ between the inertial reference systems $S$ and $S'$ must be of the type Poincar\'e-Lorentz transformation, but   
  one might bring the objection that the mass is not a Lorentz invariant as the charge is. However, this does not represent an obstacle to propose a relation of the type (40) since one must have a linear relation between these vector fields defined in $S$ and $S'$. If Poincar\'e-Lorentz transformation is assumed valid here, then, one could use the function $q(\widetilde\beta)$ as the same function used on special relativity theory $q(\widetilde\beta)=\gamma(\widetilde\beta)=(1-|\widetilde\beta|^2)^{\-1/2}$. 
\section{Conclusions and comments}
It has been shown that Maxwell-like equations for gravitational field can appear without having any relation at all with General Relativity. In fact, these type of equation could have been appeared much earlier than the formulation of the electrodynamics. Thus, it is absolutely astonishing the existing closely relation between post Newtonian gravity theory and electrodynamics theory. Several implications of Maxwell-like gravitational field formulation has been shown, based on what we already know from Electrodynamics formulation. There are several results that one need to point out: First, the energy associated to gravitational field or gravitational waves is a well defined concept. Second, linear gravitational waves appears from this extended Newtonian gravity theory, and General Relativity is not needed at all to explain their existence and their properties. Third, gravitational waves radiation appear  for any accelerated object of mass $m$, and dipole type of radiation is always expected. Using this extended Newtonian approach for gravity, it is possible to include in the body equations of motion  the dissipative effect caused by the emission of gravitational energy due to the acceleration of the body. This has been done through the gravitational radiation reaction force depending of the external force. Finally, due to this vector formulation of the linear gravity, one could say that  to unify gravity with quantum mechanics, gravity could be considered as a vector field quantity with spin one  boson as interacting quantum particle, instead of tensor quantity with spin two boson as interacting quantum particle (General Relativity),  and one could do Quantum Gravitodynamics (QGD) under this approach through (\ref{LaG}), although a concern could arisen  for not having a small parameter on this expression.  With respect the resent claimed gravitational waves detections [28], maybe it is possible to see a component of the Maxwell-like gravitational waves here, or these same waves be of Maxwell-type gravitational waves. However, these statements would require a deep careful analysis which could be made later on. 
\newpage\noindent
{\Large\bf References}\\ \\
1. I. Newton, {\it The Mathematical Principles of Natural Phylosophy}, N.Y. published by Daniel Adee, (1846).\\ \\
2. H. Goldstein, C. Poole, and J. Safko, {\it Classical Mechanics}, Third Edition, Addison Wesley, (2000).\\Ê\\
3. L.L. Helms, {\it Potential Theory}, Springer-Verlag, London, (2009).\\Ê\\
4. A. Einstein, Annalen der Phys., {\bf 49}, (1916) 769.\\Ê\\
5. M. Spivak, {\it A comprehensive Introduction to differential Geometry}, vol.1-5, Publish or Perish, Inc., (1999).\\Ê\\
6. C. M\o ller,{\it The Theory of Relativity}, Clarendon Press, Oxford, (1960)\\ 
C.W. Misner, K.S. Thorne, and J.A. Wheeler, {\it Gravitation}, W.H. Freedman and Co., (1973).\\
S. Weinberg,{\it Gravitation and Cosmology, Principles and Applications of the General Theory of Relativity}, John Wiley\& Sons, N.Y., (1972).\\Ê\\
7. B. Schulz, arXiv:1409.7977 [gr-gc], (2014).\\Ê\\
8. T. Kaluza, Sitzumgsber. Press. Akad. Wiss., Berlin (Math. Phys.), {\bf 1921}, (1921) 966.\\
O. Klein, Z. Phys., {\bf 37}, (1926) 895.\\ \\
9. M.B. Geen, J.H. Schwarz, and E. Witten, {\it Superstring Theory}, Cambridge University Press, (1987).\\
S.L. Schwebel, Int. J. Theo. Phys., {\bf 3}, (1970) 315.\\
D. Weigarten, J. Math., Phys., {\bf 18}, (1977) 165.\\
V.V. Kassandrov, Grav. Cosmol.,{\bf 1}, (1995) 216.\\
D.N. Vollick, Phys. Rev. D, {\bf 69}, (2004) 064030.\\
L.F.O. Costa and C.A.R. Herdeiro, Phys. Rev. D, {\bf 78}, (2008) 024021.\\Ê\\
10. B. Mashhoon, Phys. Lett. A, {\bf 173}, (1993) 347.\\Ê\\
11. S. Kopeikin and B. Bashhoon, Phys. Rev. D, {\bf 65}, (2002) 064025.\\Ê\\
12. R. Jantzen,, P. Carini, and D. Bini, Ann. Phys., {\bf 12}, (1995) 499.\\Ê\\
13. M.A.G. Bonilla and J.M.M. Senovilla, Phys. Rev. Lett., {\bf 11}, (1997) 783.\\Ê\\
14. S.J. Clark and R.W. Tucker, Class. Quantum Grav., {\bf 17}, (2000) 4125.\\Ê\\
15. L. Iorio and D.M. Luchesi, Class. Quantum Grav., {\bf 20}, (2003) 2477.\\Ê\\
16. L.F. Costa and C.A.R. Herderio, Phys. Rev. D, {\bf 78}, (2008) 024021.\\Ê\\
17. L.F. Costa, J. Natario and M. Zilhao, Phys. Rev. D, {\bf 93}, (2016) no.10, 104006.\\ \\
18. A. Bakopoulos and P. Kanti,, Gen. Rel. Grav., {\bf 46}. (2014) 1742.\\Ê\\
19. A. Bakopoulus and P. Kanti,, Gen. Rel. Grav., {\bf 49}, 3, (2107) 44.\\Ê\\ 
20. O. Heaviside, {\it Electromagnetic Theory, I}, "The Electrician", Printing and Publishing Co., London, (1894).\\
K.T. McDonald, Am. J. Phys. {\bf 65}, 7 (1997). \\Ê\\
21. O. Heaviside, The Electrician, {\bf 31}, (1893) 281.\\Ê\\
22. H. Thirring, Phys. Z., {\bf 19}, (1918) 204.\\
J. Lense and H. Thirring, Phys. Z., {\bf 19}, (1918) 156.\\Ê\\ 
23. K. Gr. Yaroslav, {\it Electricity, Gravity, Heat-Another Look}, Saint-Petersburg, Russia: International, (2015).\\Ê\\
24. J.D. Jackson, Classical Elecrrodynamics, John Wiely, 1975.\\ \\
25. V.S. Vladimirov, Equations of Mathematical Physics, Marcel Dekker, Inc., 1971.\\ \\
26. C. M$\not o$ller, The Theory of Relativity, Oxford at the Clarendon Press, 1952.\\
C.H. Misner, K.S. Thorne, and J.A. Wheeler, Gravitation, W.H. Freedman, 1973.\\
S. Weinberg, Gravitation and Cosmology, John Wiley \& Sons, Inc., 1972.\\
L.D. Landau and E.M. Lifshitz, The Classical Theory of Fields, Pergamon Press, 1971.\\ \\
27. G.V. L\'opez, Ann. Phys., {\bf 365}, (2017) 1.\\ 
G.V. L\'opez, J. Appl. Math. Phys., {\bf 5}, (2017) 569.\\Ê\\ 
28. B.P. Abbott(LIGO-VIRGO), Phys. Rev. Lett., {\bf 116}, (2016) 061102.\\
D. Castelvecchi and A. Witze,  Nature News, 11 Feb (2016).\\
B.P. Abbott (LIGO-VIRGO), Astrophysics, J. Lett., {\bf 818} (2016) L22.\\Ê\\
29. T.D. Padalia, Astrophysics and Space Science, {\bf 185}, (1991) 101.\\
J.H. Taylor and J.M. Weisberg, J. M. Astrophys. J., {\bf 253}, (1982) 908.\\
Lin-Sen Li,  J. Astrophysics. and Astr., {\bf 35}, (2014) 189.\\ \\
30. PDG particle data group, Particle Physics Booklet, 2018.\\ \\

\end{document}